\documentclass[prb,twocolumn,amsmath,amssymb,showpacs]{revtex4}\usepackage{graphicx}\def\beq{\begin{eqnarray}}
\def\eeq{\end{eqnarray}}
\def\beqa{\begin{eqnarray}}
\def\eeqa{\end{eqnarray}}

\begin{document}

\title{Role of dynamical non-double-occupancy excitations on the quasiparticle damping of the $t-J$ model in the 
large-$N$ limit}

\author{A. Foussats, A. Greco, and M. Bejas}

\affiliation{
{\dag} Facultad de Ciencias Exactas, Ingenier\'{\i}a y Agrimensura and Instituto de F\'{\i}sica Rosario (UNR-CONICET).
Av. Pellegrini 250-2000 Rosario-Argentina.
}

\date{\today}

\begin{abstract}

One-electron  self-energy  in the $t$-$J$ model was computed 
using a recently developed large-$N$ method based on the path integral representation for Hubbard operators.
One of the main features of the self-energy is its strong asymmetry with respect to the Fermi level, showing the spectra mostly concentrated at high negative energy.
This asymmetry is responsible for the existence of incoherent structures at high negative energy in the spectral functions.
It is shown that dynamical non-double-occupancy excitations are relevant for the behavior of the self-energy. It is difficult to understand 
the asymmetry shown by the self-energy from weak coupling treatments. 
We compare our results with others in recent literature. Finally, the possible relevance of our results for the 
recent high energy features observed in photoemission experiments is discussed.
\end{abstract}


\maketitle

It is commonly accepted that high-$T_c$ cuprates are strongly correlated systems. In these materials angle resolved photoemission spectroscopy (ARPES) experiments show interesting and still unexplained features: a) the low energy kink \cite{kink} at energy $\sim 40-70 meV$
and; b) the high energy anomalies \cite{pan06,xie07,meevasana07,graf07,zhang08} at $\sim 0.5-1eV$, known as waterfall.
For a theoretical description of these experiments it is necessary to calculate self-energy corrections on electronic correlated models as $t-J$ or Hubbard.
In spite of the progress by means of  numerical\cite{dagotto} and analytical\cite{SB} methods, the problem remains of huge interest.
Recently we have proposed a large-$N$ approach\cite{foussats02}  for the $t-J$ model which is based on the path integral representation for Hubbard operators (called PIH in what follows). 
In the PIH  method the spin index $\sigma$ is extended to a new index $p$ running from $1$ to $N$ and the perturbation is performed in powers of the small parameter $1/N$. 
It was shown that in leading order of $1/N$ (${\cal O}(1))$, which is equivalent to mean field level,  PIH results agree
with the slave boson\cite{SB} calculation. 
The large-$N$ expansion provides a controllable way for selecting and truncating Feynman diagrams. However, the results are 
more representative for the physical case $N=2$ when terms in powers of $1/N$ can be collected.
In this context, PIH can be  implemented 
beyond mean field allowing the calculation of self-energy corrections and spectral functions.\cite{bejas06,merinoPRB} 
The obtained spectral functions were compared with exact diagonalization results  finding  good agreement.\cite{bejas06,merinoPRB} 

In this paper we discuss the role of  dynamical fluctuations of the non-double-occupancy constraint on the self-energy results. 
In addition, differences between present  self-energy and that obtained from calculations based on weak coupling approaches like, for instance, random phase approximation\cite{mahan} (RPA) are discussed.
We compare also our results with those obtained by other calculations, and discuss the possible relevance of present results for the recent high
energy features observed in ARPES experiments in cuprates.

In Refs.[\onlinecite{foussats02,bejas06}] it was discussed that PIH  approach weakens collective spin fluctuations over charge fluctuations. 
Although for finite doping away from half filling the relevance of magnetism  is a matter of debate, 
for preventing possible  
objections about the influence of magnetic contributions,
we calculate for the high doping value $\delta=0.3$.  
This high doping corresponds to highly overdoped regime of cuprates where magnetic fluctuations are found to be very
weak.\cite{wakimoto07}
In addition, and for simplicity, we present results for $J=0$. For high doping, PIH  does not show strong dependence with $J$, being 
representative the results for $J=0$ (see Ref.[\onlinecite{greco08}] for discussion). 
On the other hand, no strong $J$ dependence is expected for  high 
doping values. 

Collecting  all ${\cal O}(1/N)$ contributions, the full self-energy $\Sigma({\bf k},\omega)$ (real and imaginary parts) in the square lattice 
is described in Refs.[\onlinecite{bejas06,merinoPRB}]. Herein, for convenience, we reproduce only the corresponding results for scattering rate:
\begin{eqnarray}\label{eq:SigmaT}
Im \, \Sigma_{T}({\mathbf{k}},\omega)&=& 
Im \, \Sigma_{RR}({\mathbf{k}},\omega)+2\;Im \, \Sigma_{R\lambda}({\mathbf{k}},\omega) \nonumber\\
&+&Im \, \Sigma_{\lambda\lambda}({\mathbf{k}},\omega)
\end{eqnarray}
\noindent where
\begin{eqnarray*}\label{eq:SigmaP}
Im \, \Sigma_{RR}({\mathbf{k}},\omega)&=&\frac{-1}{ N_{s}}
\sum_{{\mathbf{q}}}  \Omega^{2} \;
 Im[D_{RR}({\mathbf{q}},\omega-\varepsilon_{{k-q}})]  \nonumber\\
&& \hspace{-1cm}\times \left[n_{F}(-\varepsilon_{{k-q}}) + n_{B}(\omega-\varepsilon_{{k-q}})\right]
 ,
\end{eqnarray*}
\begin{eqnarray*}\label{eq:SigmaPP}
Im \, \Sigma_{R\lambda}({\mathbf{k}},\omega)&=& \frac{-1}{ N_{s}}
\sum_{{\mathbf{q}}}\Omega \;
Im[D_{\lambda R}({\mathbf{q}},\omega-\varepsilon_{{k-q}})]\nonumber\\
&& \hspace{-1cm}\times \left[n_{F}(-\varepsilon_{{k-q}}) + n_{B}(\omega-\varepsilon_{{k-q}})\right]
 , 
\end{eqnarray*}
\begin{eqnarray}\label{eq:SigmaPPP}
Im \, \Sigma_{\lambda\;\lambda}({\mathbf{k}},\omega)&=& \frac{-1}{ N_{s}}
\sum_{{\mathbf{q}}}
Im[D_{\lambda \lambda}({\mathbf{q}},\omega-\varepsilon_{{k-q}})]  \nonumber\\
&& \hspace{-1cm}\times \left[n_{F}(-\varepsilon_{{k-q}}) + n_{B}(\omega-\varepsilon_{{k-q}})\right]
 . 
\end{eqnarray}

In eq.(\ref{eq:SigmaP}),  $\Omega=(\varepsilon_{{k-q}}+\omega+2\mu)/2 $ and $\varepsilon_{k}=-t \delta (\cos k_{x}+\cos{k_{y}})-\mu$ is the mean field electronic band. $N_s$ is the number of sites,
$n_F$ ($n_B$) is the Fermi (Bose) factor and $\mu$ the chemical potential. 
$D_{ R  R}({\bf q},\omega)$ is the charge-charge correlation function which contains collective charge excitations.  $D_{\lambda \lambda}$ and $D_{ R  \lambda}$ correspond to the pure non-double-occupancy sector and the mixing between non-double-occupancy and charge sectors respectively. 

Starting from $\Sigma_T$ and $\Sigma_{RR}$ the spectral functions $A_T({\bf k},\omega)$ and $A_{RR}({\bf k}, \omega)$ are respectively calculated. In the calculation of the spectral functions, the real part of $\Sigma$ was numerically computed by using the Kramers-Kronig relation from 
eq.(1).
Spectral functions $A_T({\bf k},\omega)$  (solid lines) are presented in 
Fig. 1 from $\Gamma \;({\bf k}=(0,0))$ 
(panel (a)) to the Fermi vector  ${\bf k}_F$ (panel (f)) in the  $\Gamma-(\pi,\pi)$ direction (nodal direction) of the Brillouin zone (BZ).
The sharp peak near $\omega=0$ is the quasiparticle (QP) peak and defines the  QP Zhang-Rice\cite{ZR} (ZR) band of 
the $t-J$ model whose  bandwidth is reduced 
from that of the mean field band $\varepsilon_k$  due to the self-energy renormalizations. 
In addition $A_T({\bf k},\omega)$ shows incoherent spectrum (IS) which is mainly  localized at high negative energy ($\omega \sim -4t$).
Almost no IS is observed for $\omega >0$.
Similarly to Lanczos diagonalization results,\cite{horsch,dagoto,bejas06} while the QP disperses through the Fermi surface, the IS moves in opposite direction. 
Present results are  for temperature $T=0 K$, and a finite value for $T$ does not change the main conclusion. As discussed below,
self-energy contributions (Fig.3 ), responsible for the IS,  lie on an energy scale of the order of several $t$. Therefore, 
no significant changes occur for realistic values of temperature.

\begin{figure}
\begin{center}
\setlength{\unitlength}{1cm}
\includegraphics[width=9cm]{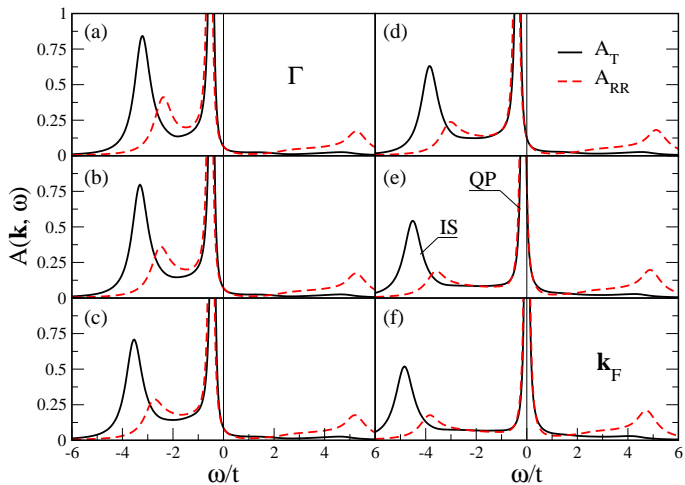}
\end{center}
\caption{(Color online) Spectral functions $A_T$ (solid lines) and $A_{RR}$ (dashed lines) from $\Gamma$ (panel (a))  to ${\bf k}_F$ 
(panel (f)) in the nodal direction for $\delta=0.3$. Energies are in units of $t$. The vertical line at $\omega=0$ marks the Fermi level.
}
\label{fig:SpecFun1}
\end{figure}

Fig.2a shows, in the main directions of the BZ,  the energy dispersion of the QP peak (solid circles) and the IS (open circles) shown 
by $A_T$. 
In panel (b) we reproduce the $t-J$ model results from Fig.1c of Ref.[\onlinecite{chinos}] obtained, also for doping $\delta=0.3$, 
using 
Gutzwiller projection variational Monte Carlo (VMC). Although both methods (PIH and VMC) are different, results are similar.
Both panels show a QP (or ZR)  band and IS at high  negative energy. None of the obtained  results show significant IS at positive energy. 
Although the IS is more dispersing in PIH than in VMC, its energy position is of the same order of magnitude in both methods. The 
QP band is somewhat flatter in PIH, if $t=0.4 eV$.\cite{chinos} For instance, the QP at $\Gamma$ is at $\omega \sim -0.4 eV $ for VMC and at $\omega \sim -0.2 eV$ for PIH. 
(In Ref.[\onlinecite{bejas06}] it was discussed that the QP bandwidth predicted by  PIH is reduced from that obtained by Lanczos).
The size of the circles scales linearly with  the spectral weight (SW).
The SW on the BZ is more homogeneous in panel (a) than in panel (b). PIH predicts, at ${\bf k}_F$, a QP weight $Z \sim 0.5$ indicating that, even for $\delta =0.30$, $\sim 50\%$ of the SW is concentrated in the IS.

\begin{figure}
\begin{center}
\setlength{\unitlength}{1cm}
\includegraphics[width=9cm]{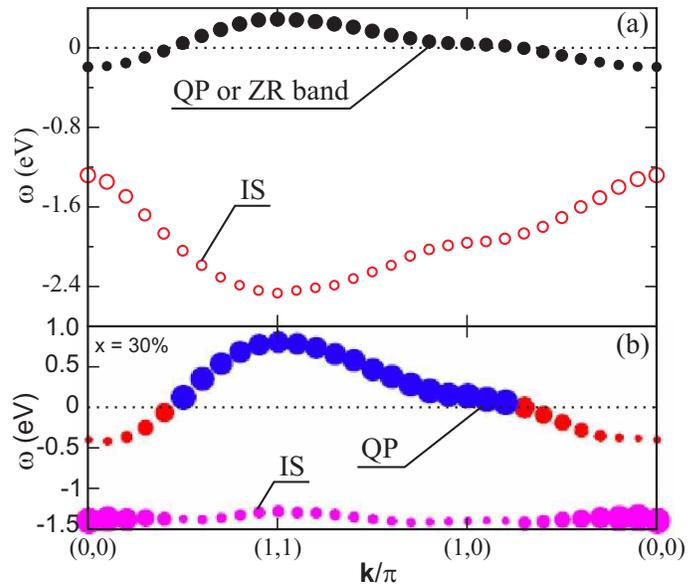}
\end{center}
\caption{(Color online) (a) Energy dispersion of the QP peak (solid circles) and the IS (open circles) shown by $A_T$ for $\delta=0.3$. (b) Results reproduced after Ref.[\onlinecite{chinos}] for a qualitative comparison with panel (a). Dotted line in both panels marks the Fermi level. 
}
\label{fig:disper}
\end{figure}

\begin{figure}
\begin{center}
\setlength{\unitlength}{1cm}
\includegraphics[width=8.5cm]{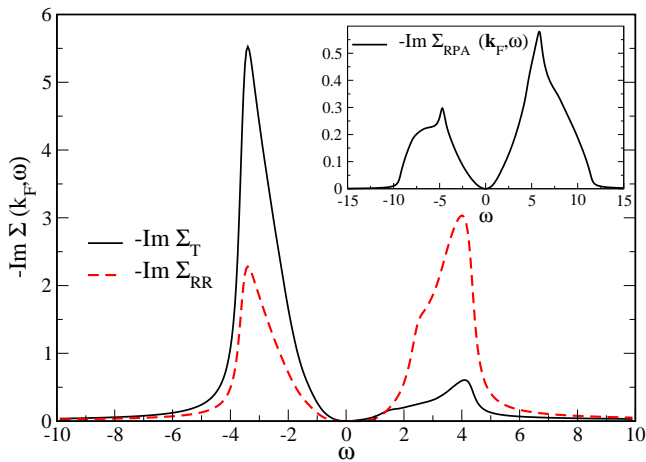}
\end{center}
\caption{(Color online) Scattering rate for $\Sigma_T$ (solid line) and $\Sigma_{RR}$ (dashed line) versus $\omega$ for $\delta=0.3$ at ${\bf k}={\bf k}_F$.
Inset, scattering rate predicted by RPA on the Hubbard model for $\delta=0.3$ at ${\bf k}_F$ (see text).}
\end{figure}

The scattering rate $-Im \Sigma_T({\bf k},\omega)$ at  ${\bf k}={\bf k}_F$ in the $\Gamma-(\pi,\pi)$ direction 
is presented in Fig. 3. 
$Im \Sigma_T$  (solid line) is very asymmetric with respect to $\omega=0$ showing most of the SW at $\omega <0$.
This asymmetric behavior is the cause of  the shape of $A_T$  in Fig.1.
This strong 
asymmetric distribution  should be interpreted as a consequence of the 
difference between addition and removal of a single electron in a correlated system. 
Recently, \cite{zemljic} using Lanczos diagonalization  in the $t-J$ model, a similar  asymmetric behavior  was also
discussed. 
It is important to notice that our scattering rate  does not show any low 
energy scale, thus it can not explain the low energy kink. If the low
energy kink is due to magnetic excitations, \cite{kinkmag} or other electronic effects, 
they are obviously not include in our approach. However it is  possible that 
the kink is due to phonons.\cite{kinkph}

For understanding which are the leading contributions responsible for the asymmetric behavior of $\Sigma_T$,  we have calculated 
$-Im \Sigma_{RR}$.  In contrast to $Im \Sigma_T$, $Im \Sigma_{RR}$ (dashed line in Fig.3) is 
very symmetric  with respect to $\omega=0$ and $A_{RR}$ (dashed lines in Fig.1) shows the IS distributed almost equally at positive and negative $\omega$.
Evidently, $\Sigma_{R \lambda}$ and $\Sigma_{\lambda \lambda}$  (eq.(\ref{eq:SigmaT}) and eq.(2)) contribute significantly, showing the relevance of non-double-occupancy excitations on the redistribution of the SW for 
leading to the final form shown by $A_T$.
In spite of the doping studied here corresponds to the highly overdoped regime of cuprates and it could expected that a weak coupling approach, like RPA, be reliable, this is an open and  controversial issue.\cite{castro04} For instance, and important for present discussion, ARPES 
experiments show, for highly overdoped samples,  high energy anomalies with similar characteristics to those in underdoped samples.\cite{pan06,meevasana07,xie07,graf07,zhang08}
It is  worth to mention that $\Sigma_{RR}$ (first line in eq.(2)) has a somewhat similar meaning to the self-energy when only charge fluctuation are considered in RPA for the Hubbard model. This later can be written as\cite{merino06}

\begin{eqnarray}\label{RPAS}
Im \Sigma_{RPA}(\textbf{k},\omega)&=&-\frac{1}{N_{s}}\sum_{\textbf{q}}
(U/2)^2 Im [\chi_{c}(\textbf{q},\omega-\xi_{k-q})]\nonumber\\
&\times&[n_{B}(\omega-\xi_{k - q}) + n_{F}(-\xi_{k - q})]\nonumber\\
\end{eqnarray} 

\noindent where $\chi_{c}$ is the RPA charge susceptibility,\cite{mahan} $U$ is the onsite Coulomb repulsion, and $\xi_{k}= -2\;t (\cos k_{x}+\cos{k_{y}}) - \mu$ is the bare tight-binding band on the square lattice.
Since in both, $Im \Sigma_{RR}$ and $Im \Sigma_{RPA}$, charge excitations are 
involved in the electron renormalizations, it is instructive to compare results from both sides. 
Inset in Fig.3 shows $-Im \Sigma_{RPA}$ at the nodal ${\bf k}_F$ for $U=4$ and $\delta=0.3$. 
Interestingly, $Im \Sigma_{RPA}$ has  a  similar shape 
to that for 
$Im \Sigma_{RR}$, i.e., $-Im \Sigma_{RPA}$ is very symmetric with  respect to $\omega=0$, leading also (not shown) to IS homogeneously distributed at both sides of the Fermi level as for the case of $A_{RR}$ (Fig.1). 
We have found this behavior for $\Sigma_{RPA}$ very stable against different conditions for $U$ and hole doping away from half-filling.
$\Sigma_{RR}$ and $\Sigma_{RPA}$ show similarities because they  can be simply interpreted in terms of fermions interacting 
with charge fluctuations. In our opinion this self-energy behavior, symmetrically distributed around $\omega=0$, can be 
expected in weak coupling.
However, results are different  when the full self-energy ($\Sigma_T$) is considered. $\Sigma_T$ is obtained in strong coupling and contains fluctuations above mean field level which are very difficult to be obtained perturbatively from usual fermions.
The  strong coupling calculation suggests that electrons interact with charge fluctuations and with excitations which represent non-double-occupancy effects. These excitations, expressed in our approach by  $D_{R \lambda}$ and $D_{\lambda \lambda}$, are  responsible of the concentration of the incoherent spectral weight at negative energy. In addition, they are dynamical  (${\bf q}$ and $\omega$ dependent) and, they can not be simply considered 
as a static enforcement of non-double-occupancy constraint as in mean field approximation.
At this point we wish to emphasize about the important role of the non-double-occupancy  constraint even for the high doping studied here. In addition to the results in Ref.[\onlinecite{chinos}] for $\delta=0.30$, Lanczos results\cite{dagoto} for $\delta=0.25$   
show also large IS at negative $\omega$. Since this behavior can be understood if the scattering rate is asymmetric with respect to $\omega=0$, 
we think that these results indicate that the overdoped $t-J$ model shows strong coupling features.  
With increasing doping, our results will be closer to those obtained using RPA. For $\delta \gtrsim  0.7$, 
$\Sigma_T$ becomes almost symmetric and, in this case, 
$\Sigma_{RR}$ approaches $\Sigma_T$ showing that $\Sigma_{R\lambda}$ and $\Sigma_{\lambda \lambda}$ have little influence.

Next, we  discuss the possible relevance of present results for the
high energy anomalies observed in ARPES experiments in  cuprates.
Momentum distribution curves (MDC) analysis of the experiment suggests 
the occurrence  of one band which is strongly renormalized near the Fermi surface. Away from the Fermi surface this band develops an 
abrupt change reappearing at 
high energy ($\sim -1eV$), given the impression of a waterfall. 
This result  supports a description in terms of renormalizations of the LDA \cite{xie07,graf07} or an uncorrelated band.\cite{markiewicz}
In contrast, energy distribution curves (EDC) analysis shows the simultaneous presence of both, low energy and high energy excitations.\cite{zhang08}
These results  suggest the occurrence  of a low energy band, associated with the ZR band of the $t-J$ model, and IS at high binding energy.\cite{greco07,chinos,zemljic}
Therefore, our results are in closer  agreement with the interpretation obtained  from the EDC
analysis. At this point it is important to remark that
it was recently discussed\cite{fink1}
that the waterfall dispersion is not an intrinsic feature but results from the suppression of the photoemission intensity 
near $\Gamma$ due to matrix elements effects (see also Ref.[\onlinecite{zhang08}] for discussion about the 
momentum and energy distribution curves dichotomy). 
Concluding, we propose that the high energy features can be described in the framework of the $t-J$ model which shows the existence of a low energy ZR band and IS at high negative energy. Additionally, we have shown that dynamical non-double-occupancy excitations are relevant for 
transferring most of the SW to high negative energy, leading to a well pronounced IS at $\omega <0$ as the observed by the experiment. In 
addition to the requirement that the IS should be mainly concentrated at negative $\omega$, its SW should be large enough to be observed. As discussed above, this condition is also satisfied by the $t-J$ model.

In summary,  we have discussed that 
dynamical non-double-occupancy effects, which are only obtained beyond 
mean field level, are responsible for a strong asymmetry  of the 
self-energy with respect to $\omega=0$.
This leads to spectral functions 
where large IS is present at high negative energy with nearly 
no signals of IS at positive $\omega$. 
It was also discussed that this picture is
very improbable to be obtained from methods which treat the electronic correlations in 
weak coupling. 
Our results show similarities with the recent 
high energy features observed by ARPES experiments in cuprates giving an additional support to the point of view that electronic correlations push cuprates to the strong coupling regime.

\noindent{\bf Acknowledgments} 
We thank Qiang-Hua Wang for valuable discussions and H. Parent
for critical reading the  manuscript.

\end{document}